\documentstyle[12pt]{article}
\newcommand{\beq}{\begin{equation}}
\newcommand{\eeq}{\end{equation}}
\newcommand{\beqa}{\begin{eqnarray}}
\newcommand{\eeqa}{\end{eqnarray}}
\newcommand{\ba}{\begin{array}}
\newcommand{\ea}{\end{array}}

\begin{document}

\begin{center}
{\Large \bf Energy Level Quasi-Crossings: \\
Accidental Degeneracies or Signature of Quantum Chaos?}
\vskip 0.5 truecm
{\bf V.R. Manfredi}$^{(1)}$ and {\bf L. Salasnich}$^{(2)}$ 
\vskip 0.5 truecm
$^{(1)}$Dipartimento di Fisica ``G. Galilei'', Universit\`a di Padova, \\
Istituto Nazionale di Fisica Nucleare, Sezione di Padova, \\
Via Marzolo 8, I-35131 Padova, Italy 
\vskip 0.3 truecm
$^{(2)}$Istituto Nazionale per la Fisica della Materia, Unit\`a di Milano, 
\\
Dipartimento di Fisica, Universit\`a di Milano, \\ 
Via Celoria 16, I-20133 Milano, Italy 
\end{center}

\vskip 1.cm 

\begin{center}
{\bf Abstract} 
\end{center}

In the field of quantum chaos, the study of 
energy levels plays an important role. The aim of this review 
paper is to critically discuss some of the main contributions 
regarding the connection between classical dynamics, 
semi-classical quantization and spectral statistics 
of energy levels. In particular, we analyze in detail degeneracies 
and quasi-crossings in the eigenvalues of quantum Hamiltonians which are 
classically non-integrable. 


\newpage

\section{Introduction}

In the last few years, many attempts have been made by 
authors working in different fields to point out some 
characteristic properties of quantal systems whose 
classical analogs are chaotic [1-4]. 

In this review paper we shall focus our attention mainly on 
the connection between quasi-degeneracies of the energy levels 
of a given quantum Hamiltonian ${\hat H}$ and the chaotic behavior of 
its classical limit. 

As stressed by Berry [5], 
if a system is without symmetries then degeneracies of the 
energy levels are considered to be accidental. 
In fact, for a typical Hamiltonian ${\hat H}$ representing a 
bound quantal system, no two of the energy levels $E$ will coincide. 
Nevertheless, if one embeds ${\hat H}$ in a 
population smoothly parameterised 
by variables ${\vec R}=(X,Y,Z,...)$ then degeneracies of 
energy levels can rise from negligible to inevitable. 
On the basis of Von Neumann and 
Wigner's famous theorem [6] for typical families ${\hat H}({\vec R})$ of real 
Hamiltonians, two parameters are necessary to produce a degeneracy, 
while if ${\hat H}({\vec R})$ is Hermitian ({\it and not real}) 
three parameters are necessary. 

A very different way of embedding a given Hamiltonian in a 
two-parameter family is to consider it as a member of a family 
labelled by a single {\it complex} parameter $C$ [7]. 
It is possible for levels to degenerate at isolated points 
in $C$-plane, even for one-dimensional problems [8]. 
But, as discussed by Berry [7], with a complex $C$ the operators 
are not Hermitians, the eigenvalues are not real, 
and at degeneracies their surfaces have a branch-point, 
rather than conical structure. 

\section{Quasi-Crossing and Chaos} 

As previously discussed, a family of generic hamiltonians 
${\hat H}(X)$ (real or Hermitian) depending on a single 
{\it real} parameter $X$ {\it will not exhibit degeneracies}: 
in the plane $E,X$ the eigenvalue curves $E_n(X)$ will not cross [6]. 
Instead, the curves will display quasi-crossings (QC) whose local 
form (which arises from a close approach to a 
degeneracy in an extended parameter space) 
is that of a pair of hyperbolas (slices of a cone or 
hypercone) [5]. Such QC are now familiar in calculations 
of the quantal spectra of families of classical 
non-integrable systems. 

If the family consists of Hamiltonians 
${\hat H}(X)$ which are not generic, 
but are special in some way, then degeneracies can be expected 
to occur. An important case is families of separable systems. 
For example, in two dimensions a particle of unitary mass 
in a rectangular box with side ratio $X$ and characteristic length 
$L$ has levels labelled by quantum numbers $m,n$ with energies 
\beq
E_{m,n} = {\hbar^2 \pi^2 \over 2 L^2}( m^2 + X^2 n^2 ) \; ,
\eeq
which can degenerate when $X^2$ is rational. 
Berry and Tabor [9] showed that 
for quantum systems whose classical analogs are integrable, 
the distribution $P(s)$ of nearest-neighbor spacings 
$s_i=({\tilde E}_{i+1}-{\tilde E}_i)$ 
of the unfolded levels ${\tilde E}_i$ follows the Poisson distribution 
\beq 
P(s)=\exp{(-s)} \; . 
\eeq 
It means that quasi-degeneracies of energy levels have the 
maximal probability to occur. Instead, for multidimensional 
systems whose classical motion is {\it quasi-integrable}, 
i.e. perturbation of integrable systems, 
Berry [5] suggested that multiple QC 
in the quantal spectrum are associated 
with classical chaos. 
In this case, the probability 
of quasi-degeneracies is zero, thus there is level repulsion, 
and $P(s)$ is quite well reproduced by 
\beq 
P(s)= {\pi \over 2} s \exp{(-{\pi \over 4}s^2)} \; , 
\eeq 
which is the so-called Wigner distribution [10]. 
\par 
Note that, the early idea of separating the spectrum of 
energy levels in regular and chaotic components 
was proposed by Percival [11]. 
Moroverer, in general, multiple QC 
{\it are not} always associated with classical chaos. 
In fact, in pseudo-integrable systems 
(rational billiards), all trajectories are confined to 
$N$-dimensional invariant surfaces (multiple-handled 
spheres rather than tori) and there is no chaos in the 
sense of exponential separation. Nevertheless, Richens 
and Berry [12] found multiple QC in the spectrum of a 
family of such systems. {\it Isolated} QC can occur even in 
integrable systems, 
for example in the double-well potential.  
\par 
The research field on quasi-crossings and their connection 
with quantum chaos is very wide. 
For reasons of space, in the next sections 
we focus our attention only on a few selected examples taken 
from molecular, nuclear and particle physics. 

\section{Molecular Spectroscopy} 

In the field of molecular spectroscopy, Marcus [13-15] studied the 
semiclassical eigenvalues of the H\`enon-Heiles Hamiltonian [16] 
\beq 
H = {1\over 2} \left(p_x^2 + p_y^2 + w_x^2x^2 + w^2_y y^2\right) + 
\lambda x (y^2+\eta x^2) \; , 
\eeq 
with coordinates $x$ and $y$, momenta $p_x$ and $p_y$, and $w_x$ and 
$w_y$ incommensurable frequencies. $\lambda$ and $\eta$ are real 
parameters. A typical classical 
trajectory is given in Fig. 1. 
On the ellipse, the potential energy equals the total energy. 
As shown by Marcus [13-15], within 
the region ABCD, the corresponding quantum mechanical wave function is 
large and oscillatory, while, outside that region, 
it dies away exponentially 
with distance. By calculating $\oint {\vec p} \cdot d{\vec r}$ 
along the curves AB and BC, setting them equal to 
$(n_x + 1/2)\hbar$ and $(n_y +1/2) \hbar$ 
respectively, semiclassical eigenvalues were calculated [13-15]. 
The agreement with the quantum mechanical eigenvalues is 
excellent.  
\par 
Marcus [13] also discussed in great detail the difference 
between a statistical and non statistical wave function: 
{\it ``A statistical (stochastic) wavefunction is described as one that 
yields averages for dynamical quantities that are approximately equal to 
their microcanonical averages at the energy: when the states are 
sufficiently dense, a microcanonical average can be computed by 
averaging over all quantum states in a small interval $(E, E + \delta 
E)$. For large enough quantum numbers, the classical microcanonical 
average could also be used for comparison. A nonstatistical wavefunction 
corresponds semiclassically to an invariant torus when the latter 
exists, and would yield, instead, an average approximately equal to the 
classical average over that torus.''} 

A statistical wavefunction could arise in the following way. One assumes 
that, in the absence of QC, the principal effect of the 
perturbation $\lambda$ is to distort the shape and extent of the region 
largely occupied by the wavefunction, but not to otherwise change it 
drastically. In that case, if one plots the eigenvalues versus some 
perturbation parameter (e.g. $\lambda$ in equation (4)) and there is no 
QC, then each state would have a nonstatistical 
eigenfunction. If, however, two such eigenvalues approach each other and 
then repel as the perturbation parameter is increased, one has a 
QC. Such QC of two energy 
levels does not yet convey a statistical character on each of the two 
wavefunctions. Rather, in the vicinity of the QC, each 
wavefunction has some of the characteristics of the two wavefunctions 
that would have arisen had the QC not occurred. A QC (see Fig. 2) 
is an analogue of an isolated classical resonance: a 
vibrational frequency, which corresponds semiclassically to a difference 
of eigenvalues, becomes nearly zero in the vicinity of the quasi 
crossing, as it does in the classical case. 

As pointed out by Marcus [13], one way of obtaining 
a statistical wavefunction is to have many overlapping QC, 
an analogue of overlapping resonances in the classical case. 
An example of multiple QC is given in Fig. 3. 
\par 
In classical mechanics, a generic Hamiltonian  
written in action-angle variable 
$({\vec J}, {\vec \theta})$ is given by 
\beq 
H({\vec J}, {\vec \theta}) = H_0({\vec J}) + \lambda V({\vec J},{\vec \theta}) 
\; , 
\eeq 
where $H_0$ is the integrable part and $\lambda V({\vec J}, {\vec \theta})$ 
a perturbation. 
A resonance occurs when there is some value ${\vec J}^*$ of ${\vec J}$, such 
that one finds 
\beq 
{\vec m} \cdot {\vec \omega}({\vec J}^*)=0 \; , 
\eeq 
where ${\vec \omega} =\partial H_0/\partial {\vec J}$. 
The resonance is an isolated resonance when one needs to consider only one 
set of integers $m_i$ for which the previous condition holds. 
In this case, one also recalls that the Hamiltonian is integrable 
(even when multiples of 
$m$ are included); a canonical transformation to new variables 
$({\vec I}, {\vec \phi})$ can be introduced such that the Hamiltonian 
is a function only of the new action variables ${\vec I}$. 
For small $\lambda$, one can make a 2nd order Taylor expansion 
about the resonant action variable to obtain an approximate Hamiltionian.  
This Hamiltonian has the form of the pendulum Hamiltonian, for 
which one finds librations and rotations, divided by the separatrix. 
Following Chirikov [17], the onset of widespread chaos begins when 
two or more resonance regions overlap, thus when two or more 
separatices cross. 
By performing the semiclassical quantization (Bohr-Sommerfeld) 
of the classical Hamiltonian, one can determine librational and 
rotational energy levels. 
The semiclassical effect of the crossing of separarices is the 
crossing of semiclassical 
energy levels that are close to the separatix. 
In actual fact, quantum mechanically one finds only quasi-crossing because 
of the previously discussed theorem of Von Neumann and Wigner 
[6] on quantum degeneracies. 

\section{Nuclear Models} 

\subsection{ZVW Model} 

Various properties of nuclear spectra can be described 
by an ensemble of random matrices, for example the Gaussian 
Orthogonal Ensemble [18]. However, realistic 
Hamiltonians of atomic nuclei contain a large regular part, 
that takes into account shell structure. 
Zirnbauer, Verbaaschot and Weidenmuller [19], using the Hamiltonian 
\beq
{\hat H}(\lambda) = {\hat H}_0 + \lambda {\hat V}_{GOE} 
\; ,
\eeq
studied the competition between order as induced 
by ${\hat H}_0$ and chaos characteristic of situations where 
the random part of the interaction ${\hat V}_{GOE}$ dominates. 
In practice, they took a diagonal ${\hat H}_0$ with equally spaced 
eigenvalues, but they stress that their conclusions are 
valid irrespective of the special choice for ${\hat H}_0$. 
In particular, they found complete mixing of eigenstates 
whenever the norm of $\lambda {\hat V}_{GOE}$ 
is comparable or greater than the norm of ${\hat H}_0$. 
Moreover, they studied the distribution of "branch points", 
defined as the complex values of $\lambda$ for which two eigenvalues 
of the Hamiltonian (7) coincide. The branch points $\lambda$ of 
${\hat H}_0$ are the solutions of the system 
$$
det(E-{\hat H}(\lambda))=0 \; , 
$$ 
\beq 
{\partial \over \partial E}det(E-{\hat H}(\lambda))=0 \; . 
\eeq 
On the real $\lambda$-axis, the branch points are observable 
as quasi-crossings of two eigenvalues of ${\hat H}(\lambda )$. 
(As discussed in the introduction, according to 
the theorem of ref. [6], 
no branch points can occur for real values of $\lambda$.) 
If $\lambda$ is large enough, one passes a large number of QC 
and, as a result, the wave functions become more and more mixed. 
The main conclusion of the author of [19] is that 
the distribution of branch points, namely the distribution of QC for real 
values of $\lambda$, is strongly peaked at $\lambda \sim 1$: 
for this value of $\lambda$ the GOE term begins to dominate 
over the regular term ${\hat H}_0$. For larger values of $\lambda$ 
the distribution 
of branch points decreases and eventually goes to zero. 

\subsection{LMG Model} 

Another example, taken from nuclear physics, is the 
three-level Lipkin-Mashow-Glick (LMG) model [20,21], 
whose quantum Hamiltonian is 
\beq 
{\hat H}=\sum_{k=0}^2\epsilon_k 
{\hat G}_{kk} - {V\over 2} \sum_{kl=0}^2 {\hat G}_{kl}^2 \; , 
\eeq 
where 
\beq 
{\hat G}_{kl}=\sum_{m=1}^M {\hat a}^+_{km} {\hat a}_{lm} 
\eeq 
are the generators of the SU(3) group. This model describes 
$M$ identical particles in three, $M$-fold degenerate, 
single-particle levels $\epsilon_i$. Like the authors of refs. [20,21], 
we assume $\epsilon_2=-\epsilon_0=\epsilon=1$, $\epsilon_1=0$, 
a vanishing interaction for particles in the same level 
and an equal interaction for particles in different levels. 
For the SU(3) model a semi-quantal Hamiltonian can be defined as 
\beq
H(p_1,p_2,q_1,q_2;M)=<q_1p_1,q_2p_2;M|{{\hat H}\over M}|q_1p_1,q_2p_2;M>,
\eeq
where $|q_1p_1,q_2p_2;M>$ is the coherent state, given by:
\beq 
|q_1p_1,q_2p_2;M>=\exp{[z_1 G_{01} + z_2 G_{02}]} |00> ,
\eeq
with:
\beq
{1\over \sqrt{2M}}(q_k+ip_k)={z_k\over \sqrt{1+z_1^*z_1+z_2^*z_2}},
\;\;\; k=1,2
\eeq
and $|00>=\Pi_{k=1}^M a^+_{0k}|0>$ is the ground state.
Here $1/M$ plays the role of the Planck constant $\hbar$. 
\par
As discussed in great detail in [20-25], 
the classical Hamiltonian can be obtained in the thermodynamic limit 
$$ 
H_{cl}=\lim_{M\to \infty} H(p_1,p_2,q_1,q_2;M) = 
$$
$$
-1+{1 \over 2}(p_1^2+q_1^2)+(p_2^2+q_2^2)+
{1\over 4} \chi [(q_1^2+q_2^2)^2 - (p_1^2+p_2^2)^2 
$$ 
\beq
-(q_1^2-p_1^2)(q_2^2-p_2^2)-4q_1q_2p_1p_2-
2(q_1^2+q_2^2-p_1^2-p_2^2)] 
\; , 
\eeq
with $\chi ={M V}/\epsilon$. The phase space has been scaled 
to give $(q_1^2+q_2^2+p_1^2+p_2^2) \leq 2$.

In order to analyze the stability of the system, we calculated  
the periodic orbits of this model using Hamilton's equation of 
the classical Hamiltonian [22]. Such equations can be written as  
\beq 
\dot{{\vec x}}=J{\vec \nabla} H_{cl}({\vec x},\chi) \; , 
\eeq 
where 
\beq 
{\vec x}=(x_1,x_2,x_3,x_4)=(q_1,q_2,p_1,p_2) \; , \;\;\; 
{\vec \nabla} = 
({\partial \over \partial q_1},{\partial \over \partial q_2}, 
{\partial \over \partial p_1},{\partial \over \partial p_2}) 
\eeq 
and $J$ is the $4\times 4$ symplectic matrix 
\beq 
J= \left( \begin{array}{cc}  
         0 & I_2\\ 
        -I_2 & 0  
\end{array}  \right) \; , 
\eeq 
where $I_2$ is the $2\times 2$ identity matrix. 
In terms of the stability matrix $S(t)$, defined 
in the usual way 
\beq
S_{ij}(t)={\partial x_i(t)\over \partial x_j(0)} \; , 
\eeq
the largest Lyapunov exponent can be written as 
\beq
\lambda({\vec x})=\lim_{t\to \infty} {1\over t} 
\log{|S(t)|} \; ,
\eeq
where $|S(t)|$ is the norm of the matrix $S(t)$. 
This matrix can be calculated by solving its equations of motion: 
\beq
{\dot S}(t)= J {\partial^2 H({\vec x})\over \partial x^2} S(t) \; , 
\eeq
with the initial conditions 
\beq
S(0)= I \; ,
\eeq 
where $I$ is the $4\times 4$ identity matrix. The calculation 
of the full set of Lyapunov exponents is related to that of the 
eigenvalues $\sigma_i$ of the matrix $S(T)$, with $T$  
the period of the periodic orbit:  
\beq
\lambda_i({\vec x})=\lim_{t\to \infty} {1\over T} 
\log{\sigma_i} \; . 
\eeq
Using the unitarity nature of $S$, a periodic orbit 
is unstable if 
\beq
Tr(S)>4 \;\;\; \hbox{or} \;\;\; Tr(S)<0 \; , 
\eeq 
and stable if 
\beq
0<Tr(S)<4 \; . 
\eeq
It is interesting to study the change in stability of periodic 
trajectories as a function of the coupling constant $\chi$. 
In Fig. 4 the ratio between the number of stable periodic orbits 
and the number of total orbits with period $T<30$ is plotted 
versus $\chi$. For the coupling constant $\chi\in (0,3]$, 
$T_{min}\simeq 3$, as shown in Ref. [22]. As can be seen, 
the sensitivity of the orbits to a small change of $\chi$ 
is quite different for different values of $\chi$, 
reflecting the transition order-chaos as the coupling constant 
increases. For the sake of completeness, 
in Fig. 5 the density $\rho(\omega)$ of the periodic orbits 
is shown as a function of the frequency $\omega = 2\pi/T$. Like 
the power spectrum, $\rho (\omega)$ presents few peaks in the 
integrable or quasi-integrable region and a quasi-continuum 
spectrum in the chaotic region.  

In order to apply the quantal criteria to our system, 
the eigenvalues of Hamiltonian (9) must be calculated. 
A natural basis can be written $|bc\rangle$, meaning $b$ particles 
in the second single-particle level, $c$ in the third and, 
of course, $M-b-c$ in the first level. 
In this way $|00\rangle$ is the ground state with all the particles 
in the lowest level. We can write the basis states as 
\beq
|bc\rangle = \left({1\over b!c!}\right)^{1/2} 
{\hat G}^b_{21}{\hat  G}^c_{31} |00\rangle \; ,
\eeq
where $(1/b!c!)^{1/2}$ is the normalizing constant. 

From the commutation relation of $G_{kl}$ we can calculate expectation 
values of ${\hat H}/M$ and thus, 
eigenvalues and eigenstates of ${\hat H}/M$. 
In this way the energy spectrum 
range is independent of the number of particles: 
\beq
\langle b' c' | {{\hat H} \over M} | b c \rangle = 
{1\over M} (-M+b+2 c) \delta_{bb'} \delta_{cc'} 
-{\chi \over 2 M^2} Q_{b'c',bc} \; ,
\eeq
where 
\beq
Q_{bc,b'c'} = 
\sqrt{b(b-1)(M-b-c+1)(M-b-c+2)} \delta_{b-2,b'}\delta_{cc'}
\eeq
$$
+\sqrt{(b+1)(b+2)(M-b-c)(M-b-c-1)} \delta_{b+2,b'}\delta_{cc'}
$$
$$
+\sqrt{c(c-1)(M-b-c+1)(M-b-c+2)} \delta_{b,b'}\delta_{c-2,c'}
$$
$$
+\sqrt{(c+1)(c+2)(M-b-c)(M-b-c-1)} \delta_{b,b'}\delta_{c+2,c'}
$$
$$
+\sqrt{(b+1)(b+2)c(c-1)} \delta_{b+2,b'}\delta_{c-2,c'}
$$
$$
+\sqrt{b(b-1)(c+1)(c+2)} \delta_{b-2,b'}\delta_{c+2,c'}
\; .
$$
The expectation values $\langle {\hat H}/M \rangle$ are real and 
symmetric. For any given number of particles $M$, we can set up the 
complete basis state, calculate the matrix elements of 
$\langle {\hat H}/M \rangle$, and then diagonalize 
$\langle {\hat H}/M \rangle$ to find its eigenvalues. 
$\langle {\hat H}/M \rangle$ connects states with $\Delta b = 
-2,0,2$ and $\Delta c = -2,0,2$ only, which 
simplifies matters. States with $b,c$ even; 
$b,c$ odd; $b$ even and $c$ odd; $b$ odd and $c$ even 
are grouped together. 
Thus $\langle {\hat H}/M \rangle$ becomes block diagonal, containing 
four blocks that can be diagonalized separately; these 
matrices are referred to as $ee$, $oo$, $oe$ and $eo$. 
When the parameter $\chi =0$, 
the Hamiltonian consists of two oscillators and there are many 
degeneracies, but for $\chi \neq 0$ these degeneracies are obviously 
broken (see Fig.6). 
For a large number of particles (semiclassical limit), 
we calculated the density of quasicrossings outside the 
degeneracy region as a function of the parameter $\chi$ 
(see Fig. 7): 
\beq
\rho (\chi ) = {\Delta N \over \Delta \chi} \; ,
\eeq
where $\Delta N$ is the number of quasicrossing in the parameter range 
$\Delta \chi =0.01$. To obtain $\Delta N$ we fixed three values 
$\chi-\Delta \chi$, $\chi$ and $\chi + \Delta \chi$ and 
imposed that 
\beq
s_i(\chi - \Delta \chi) > s_i(\chi ) \; , 
\;\;\; 
s_i(\chi + \Delta \chi) > s_i(\chi ) \; , 
\eeq
where $s_i(\chi) = E_{i+1}(\chi ) - E_i(\chi)$. 
The results (Fig. 8) show a maximum of quasicrossings 
for $\chi =2$ for all classes, in agreement with 
the transition to chaos of Fig. 9. 

In order to study the sensitivity of energy levels to small 
changes of the parameter $\chi$ we used the statistics 
$\Delta^2(E)$ defined in the usual way 
\beq
\Delta^2(E) = |E_i(\chi +\Delta\chi ) + E_i(\chi - \Delta\chi) 
-2 E_i(\chi )| \; , 
\eeq 
which measures the curvature of $E_i$ in a small range $\Delta \chi$. 
To remove the secular variation of the level density, each 
spectrum was mapped into one which has a constant level 
density by a numerical procedure described in Ref. [26]. 
Fig. 8 shows $\Delta^2(E)$ for different values 
of $\chi$; we note that the maximum value of $\Delta^2(E)$ 
corresponds to the $\chi=2$ value. 

In conclusion, in the study of the transition from order to 
chaos, there is, in agreement with the authors of [11] and [20], 
a good correspondence between the classical approach, based on the 
stability matrix and Lyapunov exponents, 
and the quantal one, based on the quasicrossing distribution 
and the $\Delta^2(E)$ statistics. 

For the sake of completeness in Fig. 9 the distribution 
$P(s)$ of spacings $s$ between adjacent levels for the 
$eo$ class (nearest-neighbor spacing distribution) has been 
calculated and compared to the Brody distribution [27]: 
\beq
P(s)=\alpha (q+1) s^q \exp{(-\alpha s^{q+1})} \; , 
\eeq
with 
\beq
\alpha = \left( \Gamma\left({q+2\over q+1}\right) \right)^{q+1} \; , 
\;\;\; 0\leq q\leq 1 \; ,
\eeq
where $\Gamma(x)$ is the factorial function. 
This distribution interpolates between the Poisson distribution 
($q=0$) and the Wigner distribution ($q=1$). 
As can be seen from Fig. 11, this statistic also confirms 
the smooth transition from the regular to the chaotic 
regime discussed in the papers [20,21]. 

\section{Particle Physics and Field Theory} 

In this section we apply a quantal analog [28] 
of the Chirikov resonance criterion [17] to study 
the suppression of classical chaos 
in the spatially homogenous SU(2) Yang-Mills-Higgs (YMH) system [29-33]. 
The quantal Chirikov criterion means that 
one applies the semiclassical quantization 
to calculate the critical value of the parameters corresponding to the
intersection of two neighboring quantal separatrices. 
\par 
The SU(2) YMH system describes a 
Higgs scalar field $\phi$, coupled to the 
Yang-Mills vector fields $A_{\beta}^{a}$, $a=1,2,3$. 
The Yang-Mills fields are the gauge fields of the SU(2) group [34]. 
The Higgs field has the familiar sombrero-shaped potential 
\beq
V(\phi )=\mu^2 |\phi|^2 + \lambda |\phi|^4 \; .
\eeq 
\par
The SU(2) YMH system is a simplified version of 
the scalar QCD, which requires the SU(3) group and $8$ Yang-Mills 
vector fields [34]. Nevertheless, the equations of motion of the SU(2) 
YMH system are complex and strongly nonlinear. 
Some simplifications can be made by working in the (2+1)-dimensional 
Minkowski space ($\beta =0,1,2$) and 
choosing spatially homogenous Yang-Mills and the Higgs fields. 
Thus, one considers the system in the region where space fluctuations of
fields are negligible compared to their time fluctuations [29-33]. 
In the gauge $A^a_0=0$ and using the real triplet representation for the
Higgs field, the Hamiltonian of the system becomes 
$$
H={1\over 2}({\dot {\vec A_1}}^2+{\dot {\vec A_2}}^2)+
{\dot {\vec \phi}}^2 
+g^2 [{1\over 2}{\vec A_1}^2 {\vec A_2}^2
-{1\over 2} ({\vec A_1} \cdot {\vec A_2})^2+
$$
\beq
+({\vec A_1}^2+{\vec A_2}^2){\vec \phi}^2
-({\vec A_1} \cdot {\vec \phi})^2 -({\vec A_2} \cdot {\vec \phi})^2]
+V({\vec \phi}),
\eeq
where ${\vec \phi}=(\phi^1,\phi^2,\phi^3)$,
${\vec A_1}=(A_1^1,A_1^2,A_1^3)$, and
${\vec A_2}=(A_2^1,A_2^2,A_2^3)$.
\par
When $\mu^2 >0$ the potential $V$ has a minimum in $|{\vec \phi}|=0$,
but for $\mu^2 <0$ the minimum is:
$$
|{\vec \phi_0}|=({-\mu^2\over 4\lambda })^{1\over 2}=v
$$
which is the non zero Higgs vacuum. This vacuum is degenerate
and after spontaneous symmetry breaking, the physical vacuum can be
chosen ${\vec \phi_0} =(0,0,v)$.
If $A_1^1=q_1$, $A_2^2=q_2$, and the other components of the
Yang-Mills fields are zero, in the Higgs vacuum the Hamiltonian of the
system is:
\beq
H={1\over 2}(p_1^2+p_2^2)
+g^2v^2(q_1^2+q_2^2)+{1\over 2}g^2 q_1^2 q_2^2 \; ,
\eeq
where $p_1={\dot q_1}$ and $p_2={\dot q_2}$. Obviously $w^2=2 g^2v^2$ is the
mass term of the Yang-Mills fields. 
\par
Classical chaos was demonstrated in a pure Yang-Mills system [29], 
i.e. in a zero Higgs vacuum. The effect of a non zero 
Higgs vacuum can be analyzed by using the quantal analog 
of the Chirikov criterion [30]. 
We introduce the action-angle variables
by the canonical transformation 
\beq
q_i=({2I_i\over \omega})^{1\over 2}\cos{\theta_i} \;, \;\;\;\;
p_i=({2I_i \omega})^{1\over 2}\sin{\theta_i} \; , \;\; i=1,2.
\eeq
The hamiltonian becomes 
\beq
H=(I_1+I_2)\omega +{1\over v^2} I_1I_2
       \cos^2{\theta_1}\cos^2{\theta_2}.
\eeq
By the new canonical transformation into slow and fast variables:
$$
A_1=I_1+I_2 \; , \;\;\; A_2=I_1-I_2 \; ,
$$
\beq
\theta_1=\chi_1+\chi_2 \; , \;\;\; \theta_2=\chi_1-\chi_2 \; ,
\eeq
$H$ can be written:
\beq 
H=A_1\omega + {1\over 4 v^2}(A_1^2-A_2^2)
       \cos^2{(\chi_1+\chi_2)}\cos^2{(\chi_1-\chi_2)}.
\eeq
We now eliminate the dependence on the angles to the order $1/v^4$ by
resonant canonical perturbation theory [34]. First we average 
on the fast variable $\chi_1$. This yields 
\beq
{1\over 2\pi}\int_{0}^{2\pi}d\chi_{1}
\cos^{2}{(\chi_{1}+\chi_{2})}\cos^{2}{(\chi_{1}-\chi_{2})} ={1\over
8}(2+\cos{4\chi_2}),
\eeq 
and
\beq
{\bar H}_{cl}=A_1\omega +{1 \over 32 v^2}(A_{1}^{2}-A_{2}^{2})
(2+\cos{4\chi_2}).
\eeq
The dependence on $\chi_2$ is now eliminated by a second canonical
transformation. The Hamilton-Jacobi equation for the perturbation part
is 
\beqa [A_{1}^{2}-({\partial S\over \partial
\chi_{2}})^{2}] (2+\cos{4\chi_{2}})=K,
\\
{\partial S\over \partial \chi_{2}}=\pm
\sqrt{ A_{1}^{2}(2+\cos{4\chi_{2}})-K\over 2+\cos{4\chi_{2}} }.
\eeqa
and thus the Hamiltonian becomes 
\beq
{\bar H}=B_{1}+{1 \over 32 v^2}K(B_{1},B_{2}),
\eeq
where 
\beq
B_{1}=A_{1}, \;\;\;\;
B_{2}={1\over 2\pi}\oint d\chi_2 {\partial S\over \partial \chi_2}.
\eeq
It appears from the structure of this equation that the motion of our
system is similar to that of a simple pendulum:
for $0<K<B_1^2$ rotational motion, for $K=B_{1}^{2}$
separatrix, and for $B_{1}^{2}<K<3B_{1}^{2}$ librational motion.
On the separatrix we have $B_{1}^{2}(2+\cos{4\chi_{2}})=K$, and:
\beq
B_2=\pm {2\over \pi}\int_{a}^{b}dx
\sqrt{ B_{1}^{2}(2+\cos{4x})-K\over 2+\cos{4x} },
\eeq
where $a=-{\pi \over 4}$, $b={\pi\over 4}$ for rotational motion, and
$a=\phi_{-}(K,B_{1})$, $b=\phi_{+}(K,B_{1})$ for librational motion, with:
\beq
\phi_{\pm}(K,B_{1})=\pm {1\over 4}\arccos ({K\over B_1^2}-2).
\eeq
The appearance of a separatrix (which is not immediately obvious in the
$(p,q)$ coordinates) accounts, as is well known, for the
stochastic layers originating near it [35]. 
This corresponds to local irregular 
behavior of the quantum spectrum; one of its manifestations is
the local shrinking of the level spacing and the tendency
to avoided crossings [28]. Note that the resonant 
perturbation theory of Hamiltonian (35) can be quite 
easily extended also to the second order (see [36]). 
\par 
The approximate Hamiltonian depends only on the
actions so that a semiclassical quantization formula
can be obtained by the Bohr-Sommerfeld
quantization rules [34]. Set $B_1=m_1\hbar$ and $B_2=m_2\hbar$, then,
up to terms of order $\hbar$, the quantum spectrum is 
\beq
E_{m_{1},m_{2}}=m_{1}\hbar\omega +{1 \over 32 v^2}K(m_{1}\hbar , m_{2}\hbar ),
\eeq
where $K$ is implicitly defined by the relation 
\beq
m_{2}\hbar =\pm {2\over \pi}\int_{a}^{b}dx
\sqrt{ (m_{1}\hbar )^{2}(2+\cos{4x})-K\over 2+\cos{4x} },
\eeq
with $a=-{\pi\over 4}$, $b={\pi\over 4}$ for $0<K<(m_{1}\hbar )^2$, and
$a=\phi_{-}(K,B_{1})$, $b=\phi_{+}(K,B_{1})$ for
$(m_{1}\hbar )^{2}<K<3(m_{1}\hbar )^{2}$. 
\par
On the separatrix, where $K=(m_1\hbar )^2$, $m_2 =\pm \alpha m_1$,
with:
\beq
\alpha ={2\over \pi}\int_{-{\pi\over 4}}^{\pi\over 4}
dx \sqrt{1+\cos{4x}\over 2+\cos{4x}}.
\eeq
\par
It is clear that for $m_1$ fixed, the function $K$, and
hence the semiclassical energy $E_{m_1,m_2}$, is a decreasing function of
the secondary quantum number $m_2$, and we have a quantum multiplet [34]. 
We can calculate the value of the coupling 
constant $1/v^2$ corresponding to the intersection of the
separatrices of two neighboring quantum multiplets 
\beq
(m_1+1)\hbar\omega +{1 \over 32 v^2}K[(m_1+1)\hbar ,\alpha (m_1+1)\hbar ]=
m_1\hbar\omega +{1 \over 32 v^2}K(m_1\hbar ,\alpha m_1\hbar ),
\eeq
and so 
\beq
{1\over v^2}={ -32 \hbar\omega \over
K[(m_1+1)\hbar ,\alpha (m_1+1)\hbar ]-K(m_1\hbar ,\alpha m_1\hbar ) }.
\eeq
In this way, we have the quantal
counterpart [28] of the method of overlapping resonances developed by
Chirikov [17]. The denominator can be evaluated by the Taylor expansion and
finally 
\beq
{1\over v^2}= \left[ { -8\omega \over
{\partial K\over \partial B_{1}}-
\alpha{\partial K \over \partial B_{2}} }
\right]_{B_1=m_1\hbar ,B_2=\alpha m_2\hbar}.
\eeq
K is implicitly defined by the relation 
\beq
F[B_1,B_2,K(B_1,B_2)]=B_2-{\pi\over 2}
\int_{-{\pi\over 4}}^{\pi\over 4}dx
\sqrt{ B_1^{2}(2+\cos{4x})-K\over 2+\cos{4x} }=0,
\eeq
or 
\beq
F(B_{1},B_{2},K)=B_{2}-\Phi (B_{1},K)=0.
\eeq
As a function of $\Phi$, $1/v^2$ can be written:
\beq
{1\over v^2}= \lim_{K\to B_1^2}
\left[ { 8\omega {\partial \Phi\over \partial K}
\over
\alpha -{\partial \Phi\over \partial B_{1}}}
\right]_{B_1=m_1\hbar},
\eeq
where 
$$
{\partial \Phi\over \partial K}=-{1\over \pi}
\int_{-{\pi\over 4}}^{\pi\over 4} dx
{ 1\over \sqrt{(2+\cos{4x})[B_1^2(2+\cos{4x})-K]} }
$$
\beq
{\partial \Phi\over \partial B_1}={2\over \pi}
\int_{-{\pi\over 4}}^{\pi\over 4}dx
\sqrt{ {B_1^2(2+\cos{4x})\over B_1^2(2+\cos{4x})-K} }.
\eeq 
The result is 
\beq
{1\over v^2}={16 \omega \over m_1\hbar} \; ,
\eeq 
where $m_1\hbar \simeq E$ (the energy of the system) and $\omega = (2 v^2
g^2 )^{1\over 2}$. 
Therefore the chaos-order transition depends on the parameter
$\lambda = v^3 g/E$: 
if $0 < \lambda < \sqrt{2}/32$, a relevant region of the
phase-space is chaotic, but if $\lambda > \sqrt{2}/32$ the system
becomes regular. This result shows that 
the value of the Higgs field in the vacuum $v$ plays an important role. 
The system is regular for large values of $v$ in agreement with numerical 
calculations [30] and the Toda Criterion of negative 
Gaussian curvature [32,33]. The Yang-Mills coupling constant $g$ has the
same role. Instead, if $v$ and $g$ are fixed there
is an order-chaos transition increasing the energy $E$.
\par 
In conclusion, we have shown that, 
for the spatially homogenous SU(2) YMH system, 
the quantum resonance criterion, 
which describes the onset of widespread chaos associated to 
semiclassical crossing between separatices of different quantum multiplets, 
gives an analytical estimation of the classical chaos-order transition 
as a function of the Higgs vacuum, the Yang-Mills coupling constant 
and the energy of the system. 

\section{Conclusions} 

It is important to stress that 
multiple quasi-crossings are {\it not} associated with classical chaos, 
but simply with the {\it onset} of chaos. Although, 
the results of Marcus are not conclusive, those of 
Weidenmuller and our own suggest this conclusion. 
Note that the Chirikov criterion gives a qualitative picture 
of the transition to chaos, as well as its quantal analog. 
The semiclassical quantization 
of fully chaotic systems must be performed 
by using a path integral technique. Instead, 
the results we have presented  
are based on the torus quantization of approximate 
classical Hamiltonians, which are obtained by means of 
a perturbative approach. 

\newpage

\section*{References} 

\begin{description} 

\item{\ 1.} M.C. Gutzwiller, {\it Chaos in Classical and Quantum Mechanics}  
(Springer-Verlag, Berlin, 1990). 

\item{\ 2.} A.M. Ozorio de Almeida, {\it Hamiltonian Systems: 
Chaos and Quantization} 
(Cambridge University Press,Cambridge, 1990). 

\item{\ 3.} M.T. Lopez-Arias, V.R. Manfredi and L. Salasnich, 
Rivista del Nuovo Cimento, vol. {\bf 17}, n. 5 (1994). 

\item{\ 4.} V.R. Manfredi and L. Salasnich, Int. J. Mod. Phys. B 
{\bf 13}, 2343 (1999). 

\item{\ 5.} M.V. Berry, in {\it Chaotic Behavior in Quantum Systems}, 
Ed. G. Casati, NATO ASI Series B: Physics 120 
(Plenum Press, New York, 1985). 

\item{\ 6.} J. Von Neumann and E. Wigner, Phys. Z. {\bf 30}, 467 (1929). 

\item{\ 7.} M.V. Berry and M. Wilkinson, 
Proc. R. Soc. Lond. A {\bf 392}, 15 (1984). 

\item{\ 8.} C.M. Bender and S.A. Orszag, {\it Advanced Mathematical Methods for 
Scientists and Engineers} (Mc Grow-Hill, New York, 1978). 

\item{\ 9.} M.V. Berry and M. Tabor, J. Phys. A {\bf 10}, 371 (1977). 

\item{\ 10.} O. Bohigas, M.J. Gannoni and C. Schmit, 
Phys. Rev. Lett. {\bf 52}, 1 (1884). 

\item{\ 11.} I.C. Percival, Adv. Chem. Phys. {\bf 36}, 1 (1977). 

\item{\ 12.} P.J. Richens and M. Berry, Physica D {\bf 1}, 495 (1981). 

\item{\ 13.} R.A. Marcus, Ann. N.Y. Acad. Sci. {\bf 357}, 169 (1980). 

\item{\ 14.} D.W. Noid, M.L. Koszykowski, M. Tabor and R.A. Marcus, 
J. Chem. Phys. {\bf 72}, 6169 (1980). 

\item{\ 15.} R. Ramaswamy and R.A. Marcus, J. Chem. Phys. {\bf 74}, 
1385 (1981). 

\item{\ 16.} M. H\`enon and C. Heiles, Astr. J. {\bf 69}, 73 (1964).  

\item{\ 17.} B.V. Chirikov, Phys. Rep. {\bf 52}, 265 (1979). 

\item{\ 18.} M.L. Mehta, {\it Random Matrices and the Statistical Theory 
of Energy levels} (Academic Press, New York, 1967). 

\item{\ 19.} H.R. Zirnbauer, J.J.M. Verbaarschot and H.A. Weidenmuller, 
Nucl. Phys. A {\bf 411}, 161 (1983).  

\item{\ 20.} R. Williams and S. Koonin, Nucl. Phys. A {\bf 391}, 72 (1982). 

\item{\ 21.} D. Meredith, S. Koonin and M. Zirnbauer, Phys. Rev. A {\bf 37}, 
3499 (1988). 
 
\item{\ 22.} V.R. Manfredi and L. Salasnich, Z. Phys. A {\bf 343}, 1 (1992). 

\item{\ 23.} L. Dematt\`e, V.R. Manfredi and L. Salasnich, 
in {\it From Classical to Quantum Chaos}, 
Ed. G.F. Dell'Antonio, S. Fantoni and V.R. Manfredi, 
SIF Conference Proceedings {\bf 41}, pp. 111 
(Editrice Compositori, Bologna, 1993). 

\item{\ 24.} V.R. Manfredi, L. Salasnich and L. Dematte, 
Phys. Rev. E {\bf 47}, 4556 (1993). 

\item{\ 25.} V.R. Manfredi and L. Salasnich, Int. J. Mod. Phys. B 
{\bf 9}, 3219 (1995). 

\item{\ 26.} V.R. Manfredi, Lett. Nuovo Cimento {\bf 40}, 135 (1984). 

\item{\ 27.} T.A. Brody, Lett. Nuovo Cimento {\bf 7}, 482 (1973). 

\item{\ 28.} S. Graffi, T. Paul, H.J. Silverstone, 
Phys. Rev. Lett. {\bf 59}, 255 (1987). 

\item{\ 29.} G. K. Savvidy: Phys. Lett. B {\bf 130}, 303 (1983). 

\item{\ 30.} L. Salasnich, Phys. Rev. D {\bf 52}, 6189 (1995). 

\item{\ 31.} L. Salasnich, Mod. Phys. Lett. A {\bf 12}, 1473 (1997). 

\item{\ 32.} L. Salasnich, Physics of Atomic Nuclei, 
{\bf 61}, 1878 (1998). 

\item{\ 33.} L. Salasnich, J. Math. Phys. {\bf 40}, 4429 (1999). 

\item{\ 34.} S. Graffi, V. R. Manfredi, L. Salasnich:
Mod. Phys. Lett. B {\bf 7}, 747 (1995). 

\item{\ 35.} A. J. Lichtenberg, M. A. Lieberman, 
{\it Regular and Stochastic Motion} (Springer-Verlag, Berlin, 1983). 

\item{\ 36.} L. Salasnich, Meccanica {\bf 33}, 397 (1998).

\end{description}

\newpage

\section*{Figure Captions}

{\bf Figure 1}: A trajectory of Hamiltonian (4) with $w_x$ and $w_y$ 
incommensurable (adapted from Ref. 13). 
\\
{\bf Figure 2}: An example of a QC of two pairs of quantum states 
of the same symmetry, in a plot of eigenvalue $E$ versus changes in the 
parameter $\lambda$ for the H\`enon-Heiles system. Also given is an actual 
crossing of states ($12,\pm 12$) (split) with ($13,\pm 5$) 
-- an allowed crossing, since these two pairs of 
states are of different symmetry 
(adapted from Ref. 13). 
\\
{\bf Figure 3}: An example of overlapping QC. Plot of eigenvalues 
{\it vs} parameter $k_{122}$ in the Hamiltonian 
$H=1/2(p_x^2+p_y^2+p_z^2+w_xx^2+w_yy^2+w_zz^2)+ 
k_{122} xy^2 + k_{133} xz^2+ k_{233} yz^2+ k_{111} x^3
+ k_{222} y^3+ k_{1122} x^2y^2 +k_{2233} y^2z^2$, 
with $w_x:w_y:w_z=2:1:1$. Only eigenvalues for eigenfunctions 
even in $y$ are plotted. The absence of hidden symmetries (producing 
actual crossings) was assumed in joining the points 
(adapted from Ref. 13). 
\\
{\bf Figure 4}: Ratio between the number of stable periodic orbits 
and the number of total periodic orbits vs $\chi$, with $T< 30$; 
$T_{min}\simeq 3$ (adapted from Ref. 22). 
\\
{\bf Figure 5}: The density of the periodic orbits in different regions: 
(a) $\chi =2$ and $E< -1$; (b) $\chi =2$ and $E>-1$; 
(c) $\chi=100$ and $E<-28$; (d) $\chi =100$ and 
$-28<E<25$ (adapted from Ref. 23). 
\\
{\bf Figure 6}: A group of $40$ energy levels as a 
function of the parameter $\chi$ for the $eo$ class 
(adapted from Ref. 23). 
\\
{\bf Figure 7}: Density of quasicrossings {\it vs} $\chi$ 
for all classes (adapted from Ref. 24). 
\\
{\bf Figure 8}: $\Delta^2(E)$ vs $E$ for different 
values of $\chi$ for the $eo$ class; 
$M=102$; (a) $\chi=0.5$, (b) $\chi=2$; 
(c) $\chi =3$, (d) $\chi=5$ (adapted from Ref. 23). 
\\
{\bf Figure 9}: $P(s)$ {\it vs} $s$ for different values of $\chi$ for 
the the $eo$ class; $M=102$; (a) $\chi=0.5$, (b) $\chi=2$; 
(c) $\chi =3$ (adapted from Ref. 24). 

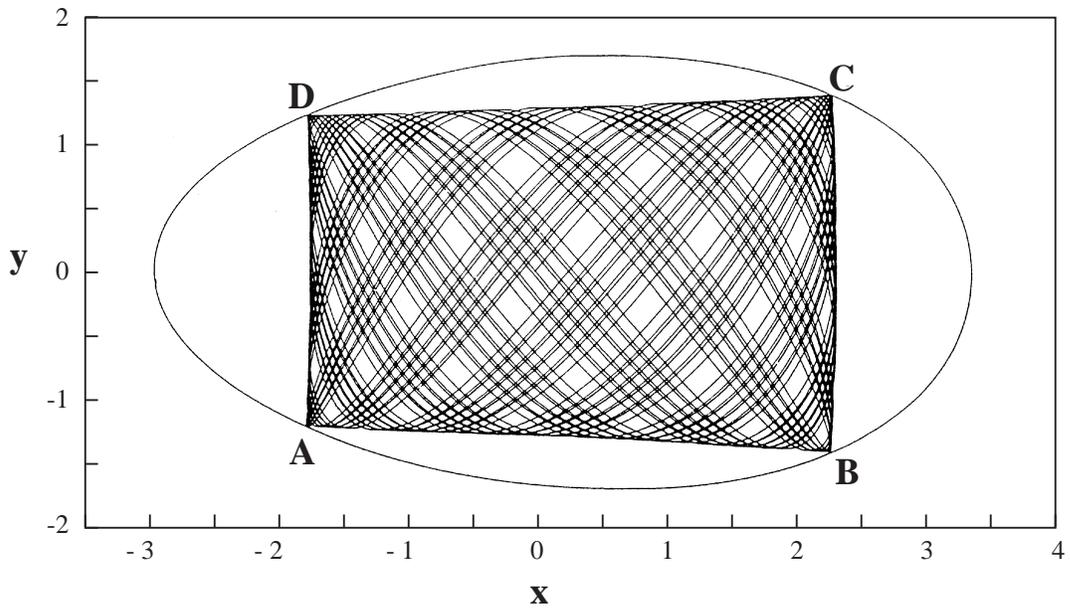
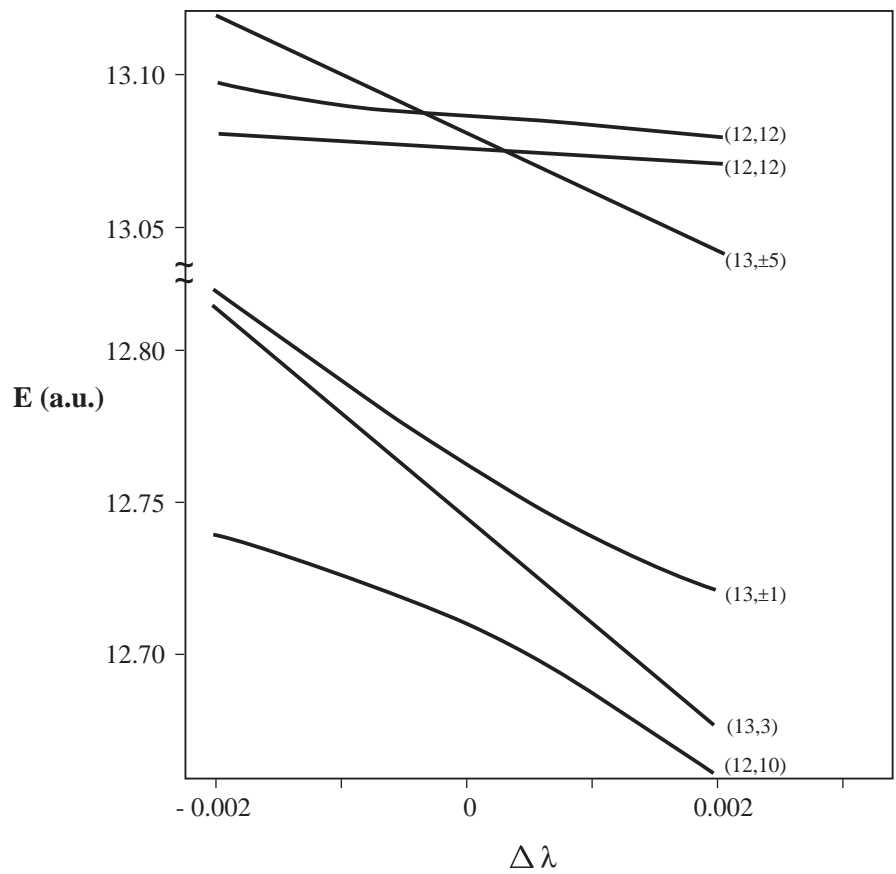
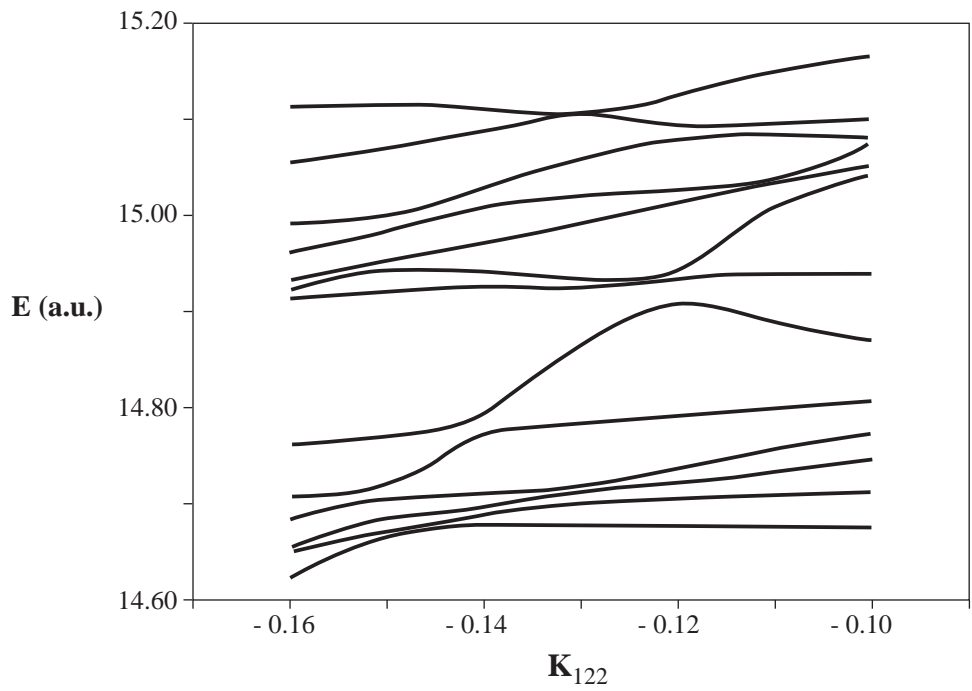
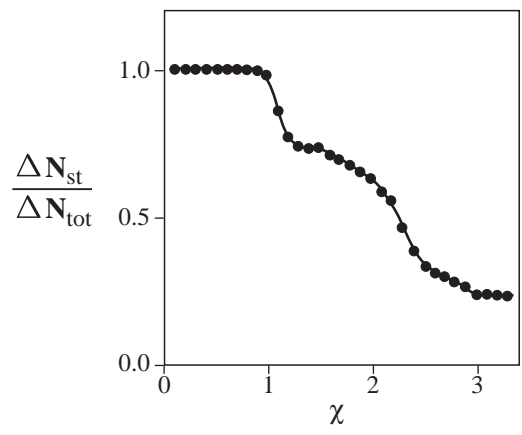
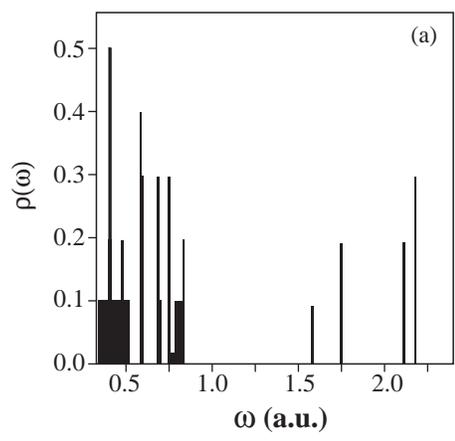
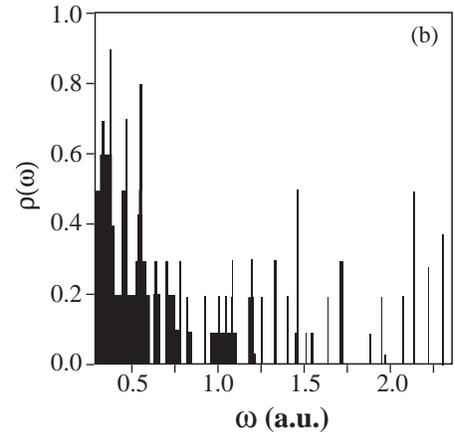
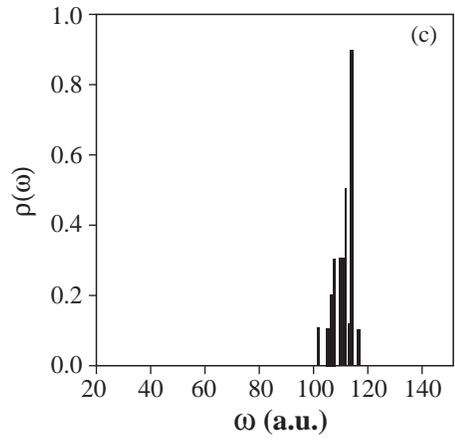
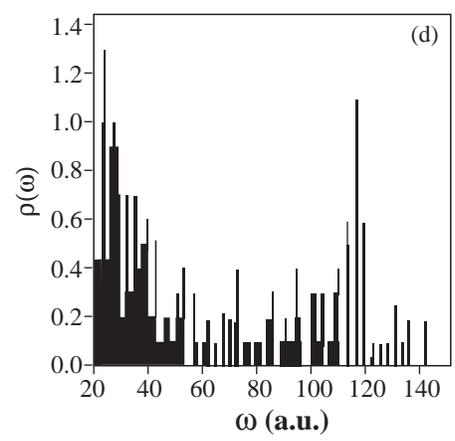
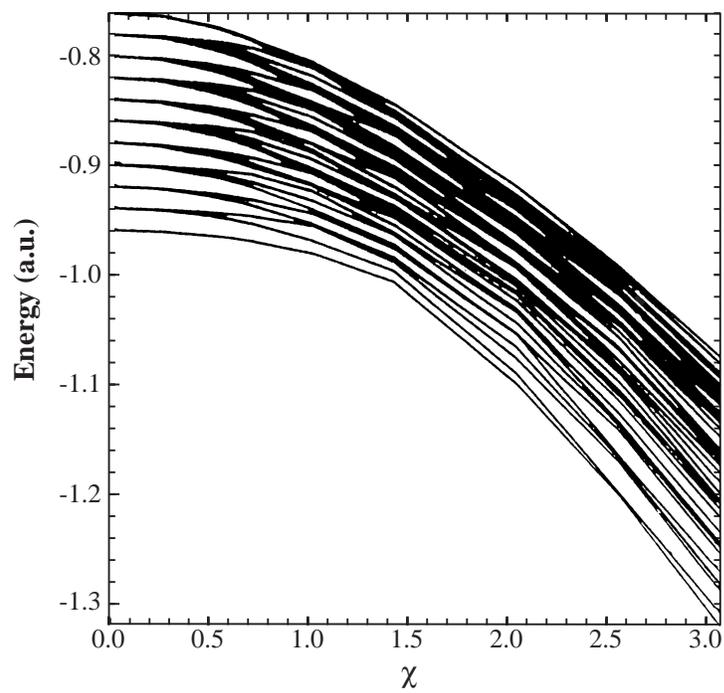
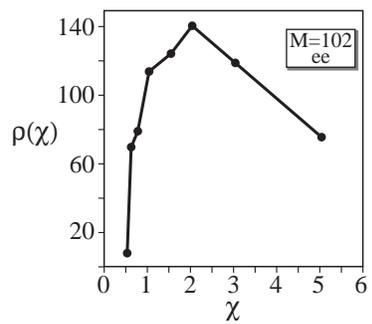
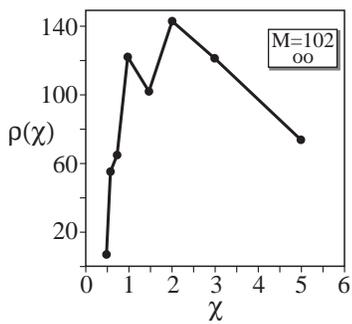
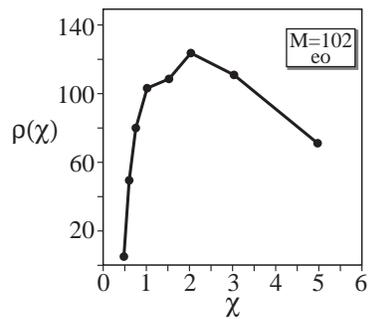
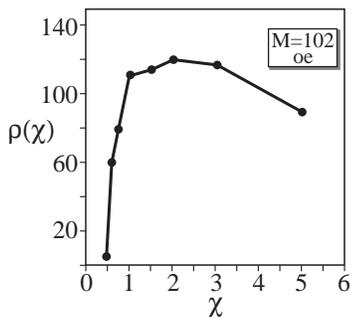
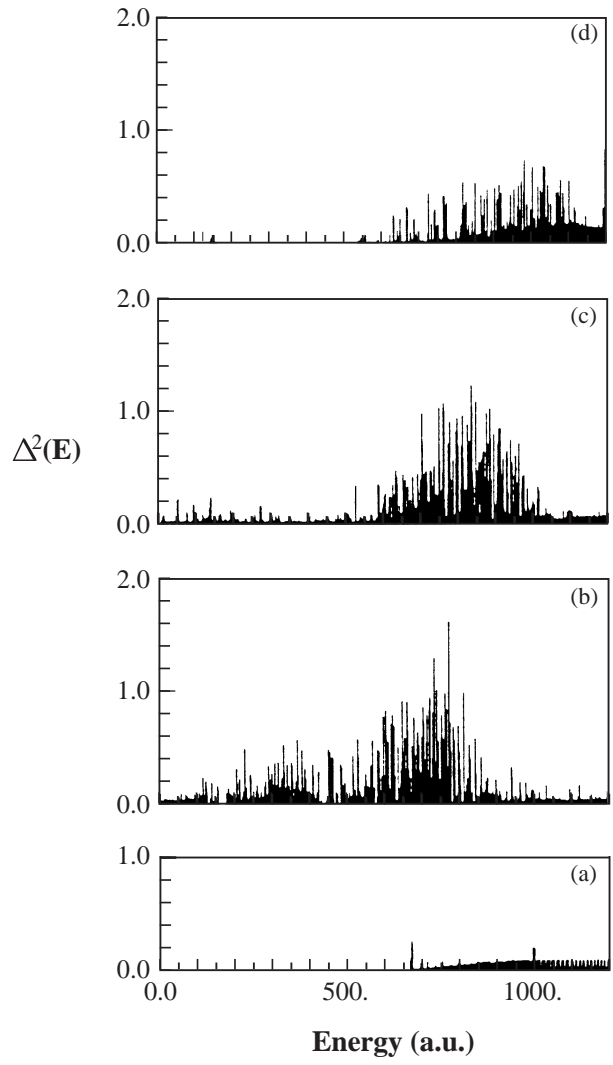
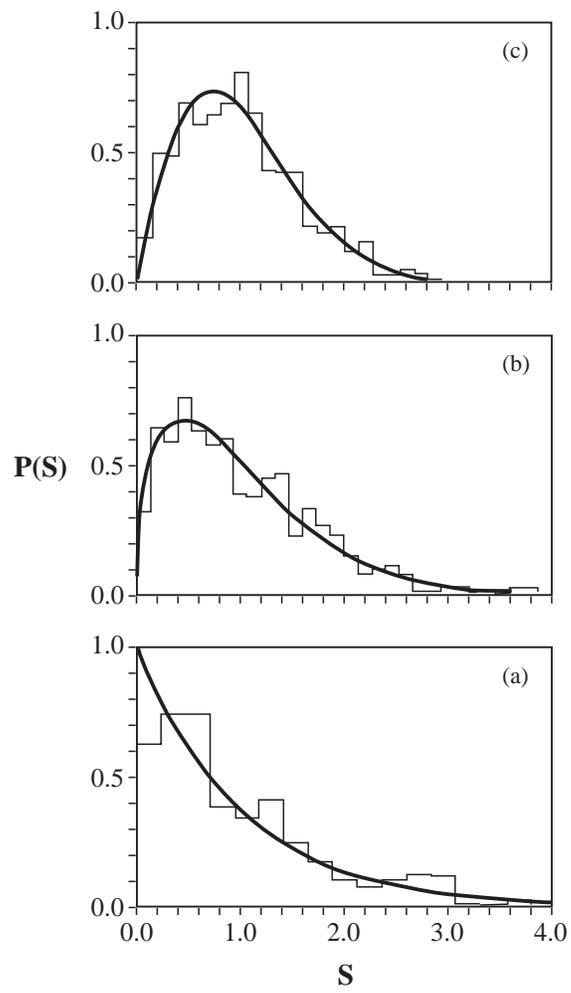
\end{document}